\documentclass[runningheads]{svmult}

\usepackage{makeidx}   % allows index generation
\usepackage{graphicx}  % standard LaTeX graphics tool
\usepackage{subeqnar}  % subnumbers individual equations
\usepackage{multicol}  % used for the two-column index
\usepackage{physprbb}  % modified textarea for proceedings,
\makeindex             % used for the subject index
\begin{document}
\title*{The Formation and Evolution of Planetary Systems: 
SIRTF Legacy Science in the VLT Era}
\toctitle{The Formation and Evolution of Planetary Systems:
SIRTF in the VLT Era}
% allows explicit linebreak for the table of content
%
%
\titlerunning{ Formation and Evolution of Planetary Systems}
% allows abbreviation of title, if the full title is too long
% to fit in the running head
%
\author{M.R. Meyer\inst{1}
\and D. Backman\inst{2}
\and S.V.W. Beckwith\inst{3}
\and T.Y. Brooke\inst{4}
\and J.M. Carpenter\inst{5}
\and M. Cohen\inst{6}
\and U. Gorti\inst{7}
\and T. Henning\inst{8}
\and L.A. Hillenbrand\inst{5}
\and D. Hines\inst{1}
\and D. Hollenbach\inst{7}
\and J. Lunine\inst{9}
\and R. Malhotra\inst{9}
\and E. Mamajek\inst{1}
\and P. Morris\inst{10}
\and J. Najita\inst{11}
\and D.L. Padgett\inst{10}
\and D. Soderblom\inst{3}
\and J. Stauffer\inst{10}
\and S.E. Strom\inst{11}
\and D. Watson\inst{12}
\and S. Weidenschilling\inst{13}
\and E. Young\inst{1}}
\authorrunning{M.R. Meyer et al.}
% if there are more than two authors
% please abbreviate author list for running head
%
%
\institute{Steward Observatory, The University of Arizona, Tucson, AZ 85721--0065 U.S.A.
\and Franklin \& Marshall College, P.O. Box 3003, Lancaster, PA  17604--3003 U.S.A. 
\and STScI, 3700 San Martin Drive, Baltimore, MD 21218 U.S.A. 
\and JPL, MS 169--237, 4800 Oak Grove Dr., Pasadena, CA  91109 
\and Caltech, Astronomy Program, MS 105--24, Pasadena, CA 91125 U.S.A  
\and Radio Astronomy Laboratory, UC--Berkeley, Berkeley, CA 94720--3411 U.S.A. 
\and NASA--Ames, MS 245--3, Mountain View, CA 94035 U.S.A. 
\and Astrophysikalisches Institut, Schillerg\"a$\beta$chen 2--3, Jena, D--07745, Germany 
\and LPL, The University of Arizona, Tucson, AZ 85721 U.S.A.
\and SIRTF Science Center, Caltech, MS 314--6, Pasadena, CA 91125 U.S.A. 
\and NOAO, P.O. Box 26732, Tucson, AZ  85726 U.S.A. 
\and University of Rochester, Rochester, NY 14627--0171 U.S.A. 
\and Planetary Science Institute, 620 N. Sixth Avenue, Tucson, AZ  85705--8331}

\maketitle              % typesets the title of the contribution

\begin{abstract}
We will utilize the sensitivity of SIRTF through the Legacy Science 
Program to carry out spectrophotometric observations of solar-type stars
aimed at (1) defining the timescales over which
terrestrial and gas giant planets are built,
from measurements diagnostic of dust/gas masses and radial distributions;
and (2) establishing the diversity of planetary
architectures and the frequency of planetesimal collisions as a function of
time through observations of circumstellar debris disks. 
Together, these observations will provide an astronomical context
for understanding whether our solar system -- and its habitable planet --
is a common or a rare circumstance.

Achieving our science goals requires measuring
precise spectral energy distributions
for a statistically robust sample capable of revealing evolutionary trends
and the diversity of system outcomes. Our targets have been selected
from two carefully assembled databases of solar-like stars:
(1) a sample located within 50 pc of the Sun 
spanning an age range from 100-3000 Myr for which a rich set of
ancillary measurements (e.g. metallicity, stellar activity, kinematics) are
available; and (2)  a selection located between 15 and 180 pc
and spanning ages from 3 to 100 Myr.
For stars at these distances SIRTF is capable of detecting stellar
photospheres with SNR $>$30 at $\lambda\leq24\mu$m for our entire sample,
as well as achieving SNR $>$5 at the photospheric limit
for over 50\% of our sample at $\lambda=70\mu$m.  Thus we will
provide a {\it complete} census of stars with excess emission
down to the level produced by the dust in our present-day solar system.

SIRTF observations obtained as part of this program will provide
a rich Legacy for follow--up observations utilizing a variety of 
facilities including the VLT.  More information concerning our program
can be found at {\tt http://gould.as.arizona.edu/feps}.
\end{abstract}

\section{Introduction}
The {\bf S}pace {\bf I}nfra{\bf R}ed {\bf T}elescope {\bf F}acility (SIRTF) is a 
key element of NASA's {\it Origins} program~\cite{ref1} .  The 85 cm cryogenic 
space telescope will be launched into an earth--trailing orbit in 2002. 
There are three science instruments on--board: IRAC, IRS, and MIPS which will provide 
imaging and spectroscopy from 3.6--160 $\mu$m for an estimated mission lifetime of $\sim$ 5 yrs. 
The {\it SIRTF Legacy Science Program} was established to provide access to large coherent datasets
as rapidly as possible in support of general observer (GO) proposals.  In addition 
to the program described here there are complementary programs to survey nearby star--forming 
(Evans et al.), the inner galactic plane (Churchwell et al.), star--formation in nearby galaxies 
(Kennicutt et al.), a wide--field extragalactic survey (Lonsdale et al.), and a 
deep pointed survey (Dickinson et al.).  For more information concerning the SIRTF Legacy Science 
Program please visit {\tt http://sirtf.caltech.edu}. 

Our modern understanding of the ubiquity of dust disks associated
with young stars began with the revelations provided by SIRTF's
ancestor IRAS ~\cite{ref2}.  Later, ISO produced a more
complete census of optically--thick disks within 200 pc 
and revealed the rich dust mineralogy and gas content of these disks
(see ~\cite{ref4} for review). 
Understanding the evolution of young circumstellar dust
and gas disks as they transition through the planet--building phase requires
the $\times$100 enhancement in sensitivity and increased
photometric accuracy offered by SIRTF at far-infrared wavlengths.
Concerning dust surrounding main sequence stars, IRAS discovered the prototypical
debris disks ~\cite{ref5} and ISO made additional limited-sample surveys 
~\cite{ref6}. Neither IRAS nor 
ISO were sensitive enough to detect dust in solar systems older than a few
hundred Myr for any but the nearest tens of stars.
SIRTF will detect orders of magnitude smaller dust masses, down to below
the mass in small grains inferred for our own present-day Kuiper Belt
(6 $\times$ 10$^{22}$ g) surrounding a solar--type star at 30 pc!

We will probe circumstellar dust properties
around a representative sample of primordial disks
(dominated by ISM grains in the process of agglomerating into planetesimals)
and debris disks (dominated by collisionally generated dust)
over the full range of dust disk optical--depths diagnostic of
the major phases of planet system formation and evolution.
Our Legacy program is designed to complement those of Guaranteed Time
Observers (GTOs) such that a direct link between disks commonly found surrounding
pre--main sequence stars $<$ 3.0 Myr old and our 4.56 Gyr old
solar system can be made.  Together, these data will help guide studies
of the formation and evolution of planetary systems undertaken with 
facilities such as the VLT.

\section{Science Strategy}
We take advantage of the efficacy of infrared
observations in revealing evidence for planetary systems
embedded in dust distributions.
In three coordinated modules we will:
1) conduct a survey of post--accretion circumstellar dust disks in order
to understand evolution of disk properties (mass and radial
structure) and dust properties (size and composition)
during the main phase of planet--building and early solar system
evolution for 150 F--G--K stars aged 3--100 Myr;
2) conduct a sensitive search for warm
molecular hydrogen in a sub--sample of 50 targets from our
dust disk survey, to constrain directly the time available
for embryonic planets to accrete gas envelopes; and 
3) trace for 150 F--G--K stars aged 100 Myr to 3 Gyr
the evolution of dust disks generated through collisions of planetesimals
and thereby infer the locations and masses of giant planets
through their action on the remnant disk.

Understanding gas--dust dynamics is crucial to our ability to
derive timescales important in planet formation and evolution.
Modules (1) and (2) have an important synergy in furthering
this understanding because
dust dynamics are controlled by gas drag rather than
radiation pressure when the gas-to-dust mass ratio is $>$10,
while it is the presence of dust that mediates gas heating
(and therefore detectability).
Module (3) investigates epochs of terrestrial and ice-giant
(Uranus- and Neptune-like) planet formation
and the subsequent dynamical evolution of planetary systems.
Our combined program will help place the formation
and evolution of our own solar system in context,
by providing the first estimates of the diversity of planetary
architectures over the full range of radii relevant for
planet formation.
Our large sample will enable us to measure the
{\it mean properties} of evolving dust disks, discover the
{\it dispersion} in evolutionary timescales, and provide a database
against which future studies can measure
how various evolutionary paths might be {\it related to stellar properties}.

\subsection{Formation of Planetary Embryos}
Our experiment begins as the disks are
making the transition from optically thick to thin, the point at which
all of the disk's mass first becomes detectable through
direct observation ~\cite{ref3}.  The goals are to:
1) constrain the initial structure and composition of
$\tau < 1$ post-accretion disks;
2) measure changes in the dust particle size distribution due to
coagulation of interstellar grains and shattering associated with
high-speed planetesimal collisions;
3) characterize the timescales over which primordial disks dissipate
and debris disks arise; and 
4) infer the presence of newly formed planets at orbital radii of 0.3-30 AU.

Photometric observations from 3.6-160$\mu$m probe temperatures (radii)
encompassing the entire system of planets in our solar system
~\cite{ref7}. 
Detailed spectrophotometry from 5.3--40 $\mu$m will
permit a search for gaps in disks caused by the dynamical
interaction of young gas giant planets and the particulate disk
from 0.2--10 AU ~\cite{ref8}.
Mid--infrared spectroscopic observations are sensitive to dust
properties including size distribution and composition which in
turn probe the physical conditions in the disk ~\cite{ref9}.
We will determine, for example, the relative importance of broad
features attributed to amorphous silicates
(ubiquitous in the ISM) compared to numerous narrow
features throughout the 5.3--40 $\mu$m region due to crystalline
dust (observed only in circumstellar environments~\cite{ref10}). In this way,
we can diagnose radial mixing in the disk because the temperature
required to anneal grains ($>$ 1500K) is substantial higher than
the inferred temperature of the emitting material ($\sim$ 300K).
Further, the shape and strength of spectroscopic features can
provide constraints on the fractional contribution of each
grain population to the total opacity; a necessary ingredient
to estimate dust mass surface densities.

\subsection{Growth of Gas Giants}
We will undertake the most
comprehensive survey to date of H$_2$ gas in post--accretion disk
systems in order to characterize its dissipation and
to place limits on the time available for giant planet formation.
We plan to survey 50 stars selected from our dust disk survey sample
at high spectral resolution (R=600) from 10--37 $\mu$m,
including both the S(0) 28 $\mu$m and S(1) 17 $\mu$m H$_2$ lines.
We focus on the post-accretion epochs from 3--100 Myr
to examine whether gas disks do indeed persist after
disk accretion onto the star has ceased ~\cite{ref11} and planetesimal agglomeration
has provided ``nucleation sites'' for gas giant planet formation~\cite{ref12}. 
We estimate that if the gas and dust temperatures are within 40K
and both are optically thin,
the dust emissivity in the continuum can
be suppressed enough to enable gas detection while
maintaining the gas temperature via collisions.
We expect to be sensitive to $>$2 $\times$ 10$^{-4}$ M$_{\odot}[H_2]$
at 70--200 K.

\subsection{Mature Solar System Evolution}
\begin{figure}
\begin{center}
\includegraphics[width=.6\textwidth]{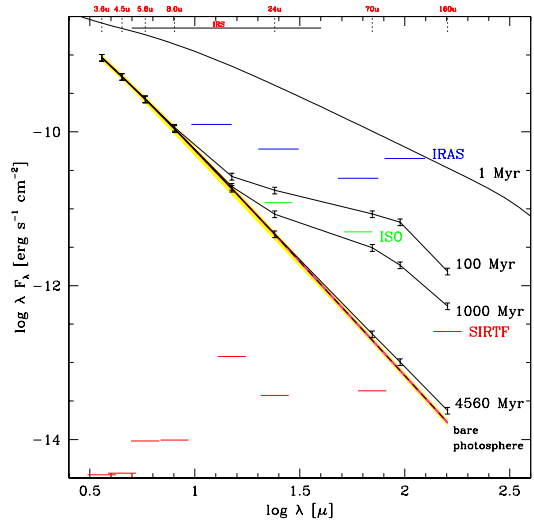}
\end{center}
\caption[]{Model SED for a hypothetical solar system based on a G2V
stellar photosphere at 30 pc, asteroid belt zodiacal dust, and Kuiper
belt dust for ages 4560, 1000, and 100 Myr
along with an optically thick disk SED characteristic
of $<$1-2 Myr old stars.  Also shown are the
IRAS, ISO, and SIRTF sensitivity limits. IRAS detected optically thick
disks out to 160 pc.  At 30 pc, ISO would have detected
this young solar system at ages of a few hundred Myr
while SIRTF will detect it at ages as old as the Sun.}
\label{sed}
\end{figure}
We complement our investigation of the initial decay of both dust and gas
signatures in first generation ``primordial'' disks with a
comprehensive study of second generation ``debris disks''.
The presence of {\it any} observable circumstellar dust around
stars older than the maximum lifetime of a primordial dust
disk (the sum of the to--be--determined gas dissipation timescale
and the characteristic Poynting--Robertson drag timescale)
provides compelling evidence not only for large reservoirs of planetesimals
colliding to produce the dust, but also for the existence of massive planetary
bodies that dynamically perturb planetesimal orbits inducing frequent
collisions~\cite{ref13}. 

We will undertake the first comprehensive survey of solar-type
stars with ages 100 Myr to 3 Gyr sensitive to dust disks comparable
to that characteristic of our own solar system throughout
its evolution from
100--300 Myr (the last phase of terrestrial and ice giant
planet--building in our solar system) through
0.3--1 Gyr (bracketing the ``late heavy bombardment'' impact peak
in our own solar system) and finally to
1.0--3.0 Gyr (examining the diversity of evolutionary paths from
early activity to mature planetary system). 
Spectroscopic observations from 5.3--40 $\mu$m enable diagnosis of gaps
caused by giant planets ~\cite{ref14} and estimates of dust size and composition which
translate directly into constraints on the mass opacity coeffients 
for the dust ~\cite{ref15} as well as Poynting-Robertson drag timescales ~\cite{ref5}. 

\section{Observing Strategy}
\subsection{Sample Characteristics}
To derive statistically meaningful results on the dust properties, we will
observe $\sim$50 stars in each of 6 logarithmically spaced age bins from
3 Myr (connecting our Legacy program to that of Evans et al.)
to 3 Gyr (beyond which there is strong
emphasis by GTO's).  Our targets span a narrow mass range (0.8-1.2~M$_\odot$)
and are proximate enough to enable a complete
census for circumstellar dust comparable to our model solar system
as a function of age (~\cite{ref17}; ~\cite{ref16}).  We will measure the stellar photosphere 
at SNR$>$30 for $\lambda\leq8\mu$m at
ages $<$ 100~Myr, and SNR$>$20 for $\lambda\leq70\mu$m at ages $>$ 100~Myr
(subject to calibration uncertainties). 
To identify gaps in the dust distribution
created by the presence of giant planets from 0.2--10 AU, we require
relative spectrophotometry with SNR$>$30 from 5.3--40 $\mu$m.
%We have estimated the photospheric flux of each star
%using extant photometry and optimized on-source SNR by considering
%backgrounds from cirrus and zodiacal emission as measured by IRAS.

For the gas evolution module, we have chosen 10 stars
for first-look observations with the high resolution mode of the IRS.
This sample spans a range of spectral type (F3--K5), age (3--100 Myr),
L$_x/L_{bol}$ ratios ($10^{-3}-10^{-5}$).
Because the line-to-continuum ratio starts to limit our
detectable H$_2$ line flux at R=600 when the continuum at 20$\mu$m is $>$100 mJy 
these sources are chosen to have optically--thin mid--infrared excess emission on
the basis of IRAS and ISO observations.
These observations will enable us to explore the limits
implied by null results and guide our choice
of follow--up observations for 40 more stars drawn from our
dust disk survey.  Our goal is a quantitative limit
on the lifetime of gas--rich disks capable of forming
giant planets.

\begin{table}
\caption{Proposed SIRTF Observations}
\begin{center}
\begin{tabular}{ccccl}
\hline\noalign{\smallskip}
\bf Instr.  & \bf \#  &  \multicolumn{1}{c}{\bf Total Time}\vline &  {\bf SNR on}& \multicolumn{1}{c|}{\bf Objectives} \\
 &{\bf stars}& \multicolumn{1}{c}{\bf inc. overhead}\vline &{\bf Photosphere}& \\\hline
\noalign{\smallskip}
\hline
\noalign{\smallskip}
IRAC  &300&40 hrs             &             & 3.6, 4.5, 5.8 and 8.0~$\mu$m\\
      &   &                   &             & photometry.                  \\
      &   &[419s/star]        &$>$30 all bands& {\bf Measure photosphere}\\
      &   &                   &                & and hot dust excess.      \\ \hline 
MIPS  &300&135 hrs                   &                  & 24, 70, and 160~$\mu$m phot.\\
      &   &                           & $>$30 at 24$\mu$m& {\bf Complete census} for dust \\
      &   &[1100-1900s/star]          & $>$5 at 70$\mu$m &  at 24 $\mu$m and most \\
      &   &                           &                  &  of sample at 70 $\mu$m. \\ \hline
IRS   &300&125 hrs                   &                         & R $\sim$ 60-120 spectra. \\
(Lo)  &   &                          & $>$30 at 5.3-14.2$\mu$m & {\bf Detailed SED}\\
      &   &[740-2020s/star]            &$>$10 at 14.2-40$\mu$m & and spectral analysis. \\ \hline 
IRS   &50&50 hrs                   &$>$3 to detect & R $\sim$ 600 spectra 10-37$\mu$m.\\
(High)&   &[720-5180s/star]  &2 $\times10^{-4}$ M$_{\odot}$ of H$_2$ & {\bf Measure H$_2$ gas mass}\\ 
      &   &                  &                                       & \& resolve dust features.\\ \hline
{\bf SUM  }  &               & {\bf 350 hrs} & & \\
\hline
\end{tabular}
\end{center}
\label{Tab1}
\end{table}
\subsection{SIRTF Data}
The SIRTF data,
in conjunction with the ancillary observations described below, will be used
to: (1) establish the contribution of the stellar photosphere to the observed
spectral energy distribution; (2) measure any excess infrared emission from
estimates of the opacity of the dust as a function of wavelength; and
(3) determine the amount, distribution, and composition of the circumstellar material
through mid--infrared spectroscopy. Our goal is to collect data capable
of realizing the fundamental limits imposed by instrument stability and
systematic calibration uncertainties.
Integration times are chosen according to each star's distance, age
and spectral type to reach uniform SNR at the photospheric
limits.
Table~1 summarizes the observations, their most basic objectives, and the
total amount of observing time requested including all overheads.

\subsection{Ancillary Data}
While SIRTF observations alone are an extremely valuable dataset, the
scientific impact of our program will be enhanced when these data are
combined with those at shorter and longer wavelengths. The ancillary data that
will be assembled and provided to the community for each star in our sample is
summarized in Table~3.
We will search for dust located at large radii and
too cold to radiate strongly in the MIPS 160 $\mu$m band by
obtaining sub--mm continuum
observations for every source in our sample. 
These observations will provide us with measurements of
(or constraints on, in the case of nondetections)
the spectral slope (F$_\nu \sim \nu^{-\alpha}$) which can be
used to constrain particle size distributions.
We also plan a limited campaign of 5 $\mu$m, 10 $\mu$m and 20 $\mu$m imaging
using the MMT, Keck, Magellan, and the VLT
for the brightest objects in our sample.  Intermediate--band photometry in
carefully chosen atmospheric windows (e.g. 5.3 and 11.7 $\mu$m) will
provide a useful check on the SIRTF calibration when tied to the same
photometric standards of Cohen et al.\ ~\cite{ref18}.
\begin{table}
\caption{Ancillary Data}
\begin{center}
\begin{tabular}{@{}llp{1.8cm}l}
\hline\noalign{\smallskip}
\bf Instrument & \bf \# stars & \bf Observing Time & \multicolumn{1}{c|}{\bf Objective}\\ \hline
Tycho/Hipparcos& 300  & Public & Proper Motions\\
Tycho/Hipparcos& 300  & Public &B$_T$, V$_T$, and H$_p$ photometry\\
2MASS          & 300  & Public &J, H, K$_S$ photometry\\
Optical Spectrographs & 300          & In hand & Spectral classification\\
Mid-Infrared Imagers &  30  & 4 runs &5-25$\mu$m imaging photometry \\
SMT/CSO/SEST & 300  & 17 weeks &sub--mm photometry  \\
\hline
\noalign{\smallskip}
\end{tabular}
\end{center}
\label{Tab2}
\end{table}
\section{The Legacy}
The combined SIRTF $+$ Ancillary Data catalog
along with specific tools for reduction, calibration,
and interpretation of data will create a rich Legacy in science
and in services provided to the community. 

\underline{First}, we will construct 3.6-160$\mu$m $+$ sub--mm
spectral energy distributions
for a representative sample of $\sim$300 F-G-K stars
aged 3 Myr - 3 Gyr within 15-160 pc of the Sun, providing a complete set
of ancillary data characterizing the properties of the stars. 

\underline{Second}, we will combine these data with model calculations
to discern the timescales for gas giant and terrestrial
planet formation in circumstellar disks, and the evolution of these systems on
Myr to Gyr timescales. 

\underline{Third}, we will provide a database of targets for follow--up
with SIRTF and other platforms.  
Our results will support GO programs with similar scientific aims
and help in selecting samples (by age, mass, metallicity, etc.) for
additional work.
Follow-up with MIPS in SED mode and with IRS at high spectral resolution
are areas in which GO's can directly exploit the Legacy database.
Our SIRTF$+$Ancillary results will also have a strong impact on future
space--based infrared surveys, as well as programs enabled by 
new ground-based facilities such as those soon available on the VLT.  
Follow--up observations with VISIR are capable of directly 
resolving the thermal disk emission from 10--20 $\mu$m ~\cite{ref19}. 
Further, high velocity resolution spectroscopic observations could 
resolve the line profiles of warm circumstellar gas 
indicating its radius of origin in a Keplerian disk ~\cite{ref20}. 
Additional studies utilizing ISAAC, CONICA, and VLTI with MIDI will 
yield further insights. 

\underline{Fourth}, due to the nature of our program 
(large sample of uniform observations) we hope to
assist the SIRTF Science Center and instrument teams in improving the photometric
accuracy of SIRTF from the initial projections of 20\% to target
values of 5 \% for the benefit of the entire astronomical community.

\underline{Fifth}, a second fundamental data product of this Legacy project
derives from the serendipitous discoveries
made as part of our primary survey.
Our program involves observing $\sim$300 fields with 38''$\times$38''
field of view to the limiting sensitivity of SIRTF at 3.6, 4.5, 5.8, \&
8 $\mu$m and these same $\sim$300 field centers with
5'$\times$5' field of view at 24, 70 \& 160 $\mu$m, over
galactic latitudes ranging from $b=20^\circ$ to $b=90^\circ$.

%INDEX%%%%%%%%%%%%%%%%%%%%%%%%%%%%%%%%%%%%%%%%%%%%%%%%%%%%%%%%%%%%%%%
% Please check with the editor of your book whether he plans to
% include a "mutual" subject index - if so, please code your entries
% in the standard syntax. For your own purposes you may print your
% "personal" index by using the following commands:
%
%\clearpage
%\addcontentsline{toc}{section}{Index}
%\flushbottom
%\printindex
%%%%%%%%%%%%%%%%%%%%%%%%%%%%%%%%%%%%%%%%%%%%%%%%%%%%%%%%%%%%%%%%%%%%%
\end{document}